\documentclass{article}
\usepackage{amssymb}

\usepackage{graphicx}
\usepackage{amsmath}
\usepackage{theorem}

\theorembodyfont{\upshape}
\newtheorem{theorem}{Theorem}

\newtheorem{lemma}[theorem]{Lemma}

\newtheorem{proposition}[theorem]{Proposition}

\input{tcilatex}

\begin{document}

\title{On the Kohn-Sham equations with periodic background potentials}
\author{E. Prodan and P. Nordlander \\
Rice University, Department of Physics-MS 61\\
6100 Main Street, Houston, Texas 77005-1892}
\maketitle

\begin{abstract}
We study the question of existence and uniqueness for the finite temperature
Kohn-Sham equations. For finite volumes, a unique soluion is shown to exists
if the effective potential satisfies a set of general conditions and the
coupling constant is smaller than a certain value. For periodic background
potentials, this value is proven to be volume independent. In this case, the
finite volume solutions are shown to converge as the thermodynamic limit is
considered. The local density approximation is shown to satisfy the general
conditions mentioned above.

\medskip

\end{abstract}

\newpage 

\section{Introduction}

For a self-interacting $N$-body quantum system with a background potential $%
u $, the finite temperature Kohn-Sham equations consist of: 
\begin{equation}
\left\{ 
\begin{array}{l}
\left( -\tfrac{1}{2}\Delta +u+n\ast v+v_{xc}\left[ n\right] \right) \eta
_{m}=\varepsilon _{m}\eta _{m} \\ 
n\left( \vec{x}\right) =\sum\nolimits_{m}\left( 1+e^{\beta \left(
\varepsilon _{m}-\mu \right) }\right) ^{-1}\left| \eta _{m}\left( \vec{x}%
\right) \right| ^{2}\text{,}
\end{array}
\right.  \label{KS1}
\end{equation}
where the value of the chemical potential is found from: 
\begin{equation}
N=\sum_{m}\left( 1+e^{\beta \left( \varepsilon _{m}-\mu \right) }\right)
^{-1}\text{.}  \label{KS2}
\end{equation}
The exchange-correlation potential $v_{xc}\left[ n\right] $ may have a
complicated functional dependence on the particle density. However, it acts
on the Kohn-Sham orbitals as a regular potential, by multiplying the
orbitals with a function. \cite{Eschrig}

These equations determine the density of particles at the thermodynamic
equilibrium of the system. The zero temperature formalism proposed by
Hohenberg and Kohn \cite{Hohenberg64}, now known as the Density Functional
Theory, can be extended to finite temperatures if, instead of the ground
state energy, one considers the grand canonical potential.\cite{Mermin65} In
this case, one can show that the grand canonical potential is a functional
of the particle density. This functional achieves its minimum for the
equilibrium particle density. Eqs. (\ref{KS1}, \ref{KS2}) represents the
Euler-Lagrange equations associated with this functional.\cite
{KohnSham65,Stoitsov88} In contrast to the zero temperature case, the finite
temperature Kohn-Sham equations involve an infinite set of orbitals. In
practical applications however, one only has to consider the orbitals with
energies up to $\mu +k_{B}T$. Above this limit, the contribution to the
density of particles Eq. (\ref{KS2}) becomes negligible. The finite
temperature formalism has a major advantage over the zero temperature
formalism in that it avoids the problem associated with the degeneracy of
the last occupied energy level.

While most of the work on these equations has been concerned with rigorous
derivations or finding better and better approximation of the
exchange-correlation potential, little it is known about their solutions. 
\cite{Bokanowski00,Kaiser99} Computational physicists often assume that the
Kohn-Sham equations have a unique solution. The same assumption has been
made for long time for models like Hartree or Hartree-Fock. Despite many
efforts however, the uniqueness is still an open problem for these models 
\cite{Lions98} and only little advance \cite{Catto01,ProdanHI,ProdanHII} has
been made in this direction. Moreover, the symmetry breaking within these
models \cite{ProdanHIII,Lieb94} shows that the uniqueness can be a very
delicate problem.

The first goal of this paper is to search for those conditions on the
effective potential that guarantee the existence and uniqueness of a
solution for the Kohn-Sham equations on a finite volume. The conditions will
be formulated at an abstract level, without making references to any
explicit expression or approximation of the exchange correlation potential.
We will show in the last section that these abstract conditions are general
enough to include the local density approximation.

The Kohn-Sham equations are derived from the Kohn-Hohenberg \cite
{Hohenberg64}\ functional. This functional is based on the assumption that
the density of particles is $v$-representable. Despite many efforts, the $v$%
-representability problem has not been yet rigorously solved for infinite
volume. A functional that does not require $v$-representability has been
proposed by Lieb.\cite{Lieb85} Unfortunately, this new functional does not
automatically lead to the Kohn-Sham equations. It was shown however that,
for finite volumes, the Kohn-Hohenberg and Lieb functionals coincide.\cite
{Eschrig} The relation between the two functionals is not yet completely
understood for infinite volume. The thermodynamic limit of the Kohn-Sham
equations can therefore be considered as a fundamental issue in density
functional theory. We believe that the unsolved part of the $v$%
-representability problem can be avoided by studying the thermodynamic limit
of the finite volume Kohn-Sham equations instead of considering the infinite
volume.

The second goal of this paper is to find the general conditions on the
effective potential which guarantee that the finite volume solutions of the
Kohn-Sham equations have a well defined thermodynamic limit. Our solution to
this problem applies only for periodic background potentials.

Within the local density approximation, partial results on the Kohn-Sham
equations have been reported in Ref. \cite{ProdanHI}. The methods developed
there could not be used for realistic exchange-correlations potentials. The
main obstacle was the low density behavior of realistic
exchange-correlations potentials. In the last section of this paper we will
show how this problem can be solved. We will show that the local density
approximation of the exchange-correlation potential satisfies the abstract
conditions mentioned above and a unique solution exists.

\section{The fixed point approach}

We assume in the following that the background potential comes from the
interaction of the particles with background charges of opposite sign, which
are considered fixed and given. In this case, the Kohn-Sham equations take
the following form: 
\begin{equation}
\left\{ 
\begin{array}{l}
\left( -\tfrac{1}{2}\Delta +\left( n-n_{0}\right) \ast v+v_{xc}\left[ n%
\right] \right) \eta _{m}=\varepsilon _{m}\eta _{m} \\ 
n\left( \vec{x}\right) =\sum_{m}\left( 1+e^{\beta \left( \varepsilon
_{m}-\mu \right) }\right) ^{-1}\left| \eta _{m}\left( \vec{x}\right) \right|
^{2}\text{.}
\end{array}
\right.
\end{equation}
We also consider a system which is charge neutral: 
\begin{equation}
\int_{vol}\left( n\left( \vec{x}\right) -n_{0}\left( \vec{x}\right) \right) d%
\vec{x}=0\text{.}  \label{Neutrality}
\end{equation}
The chemical potential must be adjusted such that the above condition is
satisfied. The neutrality condition will play an essential role when long
range interactions are considered and it will help us to improve our
previous estimates on the Hartree potential.\cite{ProdanHII} We borrowed the
idea from the homogeneous electron gas where it is known that the neutrality
condition is essential when the thermodynamic limit is considered.\cite
{Mahan90book} Also, the neutrality or partial neutrality condition play an
essential role in the Hartree-Fock model when applied to Coulomb systems. 
\cite{Lieb77,Lions87} The neutrality condition will be further discussed at
the end of Section 3.

Let us denote the Kohn-Sham Hamiltonian and the effective potential by: 
\begin{eqnarray}
H_{n} &=&-\tfrac{1}{2}\Delta +\lambda V\left[ n\right] \\
V\left[ n\right] &=&\left( n-n_{0}\right) \ast v+v_{xc}\left[ n\right] \text{%
,}  \notag
\end{eqnarray}
and the Fermi-Dirac distribution by: 
\begin{equation}
\Phi _{FD}\left( t\right) =\left( 1+e^{\beta t}\right) ^{-1}\text{.}
\end{equation}
For finite volumes, we now formulate the Kohn-Sham equations as a fixed
point problem. In this case, $\Delta $ is the Laplace operator over the
volume $vol$ constrained by various boundary conditions.

\begin{theorem}[The fixed point approach]
Let $S^{N}\subset L^{1}\left( vol\right) $ be defined as: 
\begin{equation}
S^{N}=\left\{ n\in L^{1}\left( vol\right) ,\text{ }\left\| n\right\|
_{L^{1}\left( vol\right) }=N\right\} \text{.}
\end{equation}
Suppose that for $n\in S^{N}$ and $a>0$ the following condition is
satisfied: 
\begin{equation}
\left\| V\left[ n\right] \left( -\tfrac{1}{2}\Delta +a\right) ^{-1}\right\|
\leqslant \gamma _{a}\text{,}
\end{equation}
where $\gamma _{a}$ may depend on $N$. Then the map: 
\begin{eqnarray}
T &:&S^{N}\rightarrow S^{N} \\
S^{N} &\ni &n\rightarrow T[n]\left( \vec{x}\right) =\Phi _{FD}\left(
H_{n}-\mu _{n}\right) \left( \vec{x},\vec{x}\right)  \notag
\end{eqnarray}
is well defined. Here, $\mu _{n}$ represents the unique solution of the
equation: 
\begin{equation}
N=Tr\,\Phi _{FD}\left( H_{n}-\mu _{n}\right) \text{.}  \label{miuEq}
\end{equation}
Moreover, the fixed points of the map $T$ generates all the solutions of the
Kohn-Sham equations.
\end{theorem}

Let us prove first the following result which will be used many times in the
following.

\begin{lemma}
Let $H_{0}$ be a self-adjoint, bounded from below Hamiltonian and suppose
that $\exp \left( -H_{0}\right) $ is of trace class. Let $V$ be a
self-adjoint potential such that, for $a>\left| \inf \sigma \left(
H_{0}\right) \right| $: 
\begin{equation}
\left\| V\left( H_{0}+a\right) ^{-1}\right\| \leqslant \gamma _{a}\text{.}
\end{equation}
Then $H=H_{0}+\lambda V$ is self-adjoint and: 
\begin{eqnarray}
&&Tr\phi _{j,\mu }\left[ \left( 1+\lambda \gamma _{a}\right) H_{0}+\lambda
a\gamma _{a}\right] \\
&\leqslant &Tr\,\left( 1+e^{\beta \left( H-\mu \right) }\right) ^{-j}  \notag
\\
&\leqslant &e^{j\beta \left( \mu +\lambda a\gamma _{a}\right)
}Tr\,e^{-j\beta \left( 1-\lambda \gamma _{a}\right) H_{0}}\text{,}  \notag
\end{eqnarray}
where $\phi _{j,\mu }$, $j=1$, $2$,..., are monotone decreasing, convex
functions on $\left[ \epsilon _{0},\infty \right) $, $\epsilon _{0}=\inf
\sigma \left( H\right) $, such that: 
\begin{equation}
\phi _{j,\mu }\left( t\right) \leqslant \left( 1+e^{\beta \left( t-\mu
\right) }\right) ^{-j}\text{.}
\end{equation}
All functions $\phi _{j,\mu }\left( t\right) $ can be chosen such that $%
\lim_{\mu \rightarrow \infty }\phi _{j,\mu }\left( t\right) =1$ and $\phi
_{j,\mu }\left( t\right) >0$.
\end{lemma}

\textit{Proof}. We will use classical techniques from Ref. \cite{SimonTr}.
We start with the first inequality. Let $\left\{ \eta _{m}^{0},\varepsilon
_{m}^{0}\right\} _{m}$ be the eigenvectors and the corresponding eigenvalues
of $H_{0}$. $\eta _{m}^{0}\in \mathcal{D}\left( H\right) $ and we can write: 
\begin{eqnarray}
&&Tr\left( 1+e^{\beta \left( H-\mu \right) }\right) ^{-j} \\
&\geqslant &Tr\phi _{j,\mu }\left( H\right) \geqslant \sum\nolimits_{m}\phi
_{j,\mu }\left( \left\langle \eta _{m}^{0},H\eta _{m}^{0}\right\rangle
\right)  \notag \\
&=&\sum\nolimits_{m}\phi _{j,\mu }\left( \left( \varepsilon
_{m}^{0}+a\right) \left\langle \eta _{m}^{0},\left( I+\lambda V\left(
H_{0}+a\right) ^{-1}\right) \eta _{m}^{0}\right\rangle -a\right)  \notag \\
&\geqslant &\sum\nolimits_{m}\phi _{j,\mu }\left( \left( 1+\lambda \gamma
_{a}\right) \varepsilon _{m}^{0}+\lambda a\gamma _{a}\right)  \notag \\
&=&Tr\phi _{j,\mu }\left( \left( 1+\lambda \gamma _{a}\right) H_{0}+\lambda
a\gamma _{a}\right) \text{.}  \notag
\end{eqnarray}
For the second inequality, let $\left\{ \eta _{m}\text{,}\varepsilon
_{m}\right\} _{m}$ be the eigenvectors and the corresponding eigenvalues of $%
H$. Then: 
\begin{eqnarray}
&&e^{j\beta \left( a+\mu \right) }Tr\,e^{-j\beta \left( 1-\lambda \gamma
_{a}\right) \left( H_{0}+a\right) } \\
&\geqslant &e^{j\beta \left( a+\mu \right) }\sum\nolimits_{m}e^{-j\beta
\left( 1-\lambda \gamma _{a}\right) \left\langle \eta _{m},\left(
H_{0}+a\right) \eta _{m}\right\rangle }  \notag \\
&=&e^{j\beta \left( a+\mu \right) }\sum\nolimits_{m}e^{-j\beta \left(
1-\lambda \gamma _{a}\right) \left( \varepsilon _{m}+a\right) \left\langle
\eta _{m},\left( I+\lambda V\left( H_{0}+a\right) ^{-1}\right) ^{-1}\eta
_{m}\right\rangle }  \notag \\
&\geqslant &e^{j\beta \left( a+\mu \right) }\sum\nolimits_{m}e^{-j\beta
\left( 1-\lambda \gamma _{a}\right) \left( \varepsilon _{m}+a\right) /\left(
1-\lambda \gamma _{a}\right) }  \notag \\
&\geqslant &Tr\,\left( 1+e^{\beta \left( H-\mu \right) }\right) ^{-j}\text{.}%
\blacksquare  \notag
\end{eqnarray}

\textit{Proof of Theorem 1}. We need to show that Eq. (\ref{miuEq}) has a
unique solution for all $n\in S^{N}$. From the previous Lemma, 
\begin{eqnarray}
&&\dfrac{d}{d\mu }Tr\,\Phi _{FD}\left( H_{n}-\mu \right) \\
&=&\beta e^{-\beta \mu }Tr\,\left( 1+e^{\beta \left( H_{n}-\mu \right)
}\right) ^{-2}  \notag \\
&\geqslant &\beta e^{-\beta \mu }Tr\phi _{2,\mu }\left( -\tfrac{1}{2}\left(
1+\lambda \gamma _{a}\right) \Delta +\lambda a\gamma _{a}\right) \text{.} 
\notag
\end{eqnarray}
The last term is strictly positive when we choose $\phi _{2,\mu }>0$. Also,
from the previous Lemma, it follows that the above derivative is finite.
Then the right hand side of Eq. (\ref{miuEq}) is a strictly monotone,
continuous function of $\mu $. As we already mentioned, $\phi _{j,\mu }$ can
be chosen such that $\phi _{j,\mu }\rightarrow 1$ as $\mu $ goes to $+\infty 
$. Then one can see from 
\begin{eqnarray}
&&Tr\phi _{1,\mu }\left[ -\tfrac{1}{2}\left( 1+\lambda \gamma _{a}\right)
\Delta +\lambda a\gamma _{a}\right] \\
&\leqslant &Tr\,\Phi _{FD}\left( H_{n}-\mu \right)  \notag \\
&\leqslant &e^{\beta \left( \mu +\lambda a\gamma _{a}\right) }Tr\,e^{\beta
/2\left( 1-\lambda \gamma _{a}\right) \Delta }\text{,}  \notag
\end{eqnarray}
that, as $\mu $ is varied from $-\infty $ to $+\infty $, the right hand side
of Eq. (\ref{miuEq}) varies from $0$ to $+\infty $. This, combined with the
strict monotonicity and continuity, allows us to conclude that Eq. (\ref
{miuEq}) has a unique solution. Using the eigenvectors and the eigenvalues
of $H_{n}$, 
\begin{equation}
\left( -\tfrac{1}{2}\Delta +\left( n-n_{0}\right) \ast v+v_{xc}\left[ n%
\right] \right) \eta _{m}=\varepsilon _{m}\eta _{m}\text{,}  \label{Egval}
\end{equation}
the fixed point equation for $T$ reduces to 
\begin{equation}
n\left( \vec{x}\right) =\sum_{m}\left( 1+e^{\beta \left( \varepsilon
_{m}-\mu _{n}\right) }\right) ^{-1}\left| \eta _{m}\left( \vec{x}\right)
\right| ^{2}\text{.}  \label{den}
\end{equation}
Eqs. (\ref{Egval}) and (\ref{den}) represent exactly the Kohn-Sham equations.%
$\blacksquare $

The fixed points of the map $T$ generates all the solutions of the Kohn-Sham
equations because $S^{N}$ is the largest set where these solutions can be
found. Another important observation is that there exists an upper and lower
limit on the chemical potential, limits which may depend in general on the
number of particles. This can be seen from: 
\begin{eqnarray}
&&Tr\phi _{1,\mu _{n}}\left[ -\tfrac{1}{2}\left( 1+\lambda \gamma
_{a}\right) \Delta +\lambda a\gamma _{a}\right]  \label{miulimits} \\
&\leqslant &N\leqslant e^{\beta \left( \mu _{n}+\lambda a\gamma _{a}\right)
}Tr\,e^{\beta /2\left( 1-\lambda \gamma _{a}\right) \Delta }\text{.}  \notag
\end{eqnarray}
For periodic background potentials however, we will show that this limits
are independent of the number of particles.

\section{The Kohn-Sham equations with periodic background potentials}

In the following we consider a background charge distribution which is
periodic with respect to a lattice $\Gamma $, 
\begin{equation}
\Gamma =\left\{ \vec{x}\in R^{3}\text{, }\vec{x}=\sum_{i=1}^{3}n^{i}\vec{%
\delta}_{i}\text{, }n^{i}\in Z\right\} \text{,}
\end{equation}
i.e. $n_{0}\left( \vec{x}+\vec{R}\right) =n_{0}\left( \vec{x}\right) $
almost everywhere when $\vec{R}\in \Gamma $. $\vec{\delta}_{i}$, $i=%
\overline{1,3}$, represent three linearly independent vectors in $R^{3}$.
Let us also consider a finite crystal confined in the volume: 
\begin{equation}
V=\left\{ \vec{x}\in R^{3}\text{, }\vec{x}=\sum_{i=1}^{3}\alpha ^{i}\vec{%
\delta}_{i}\text{, }0\leqslant \alpha ^{i}\leqslant K\right\} \text{,}
\end{equation}
where $K$ is a positive integer. We denote the crystal's unit cell by: 
\begin{equation}
cell=\left\{ \vec{x}\in R^{3}\text{, }\vec{x}=\sum_{i=1}^{3}\alpha ^{i}\vec{%
\delta}_{i}\text{, }0\leqslant \alpha ^{i}\leqslant 1\right\} \text{.}
\end{equation}
Thus, the crystal is formed from $K^{3}$ unit cells. We impose periodic
boundary conditions and we also allow the particles on opposite faces of the
crystal to interact to each other. The resulting problem is that of
particles trapped on a torus $\mathcal{T}$ obtained by connecting the
opposite faces of the crystal. Any point from $R^{3}$ can be viewed as a
point of the torus. The kinetic term of the Kohn-Sham Hamiltonian is given
by $-\frac{1}{2}\Delta $, where $\Delta $ represents the Laplace operator
over the torus $\mathcal{T}$. We will assume that the particles interact via
a two-body potential which depends only on the distance between particles: 
\begin{equation}
v\left( \vec{x},\vec{y}\right) =v\left( \left| \vec{x},\vec{y}\right|
\right) \text{ , \ }\vec{x}\text{, }\vec{y}\in \mathcal{T}\text{,}
\end{equation}
where $\left| \cdot ,\cdot \right| $ denotes the distance on the torus. In
this case, the potential generated by the background charge is $\Gamma $%
-periodic. Indeed, for $\vec{R}\in \Gamma $: 
\begin{eqnarray}
\int_{\mathcal{T}}v\left( \left| \vec{x}+\vec{R},\vec{y}\right| \right)
n_{0}\left( \vec{y}\right) d\vec{y} &=&\int_{\mathcal{T}}v\left( \left| \vec{%
x},\vec{y}-\vec{R}\right| \right) n_{0}\left( \vec{y}\right) d\vec{y} \\
&=&\int_{\mathcal{T}}v\left( \left| \vec{x},\vec{y}\right| \right)
n_{0}\left( \vec{y}+\vec{R}\right) d\vec{y}  \notag \\
&=&\int_{\mathcal{T}}v\left( \left| \vec{x},\vec{y}\right| \right)
n_{0}\left( \vec{y}\right) d\vec{y}\text{,}  \notag
\end{eqnarray}
where we also used that the measure $d\vec{y}$ is invariant at translations.
We will assume that the exchange-correlation potential is $\Gamma $-periodic
when the density of particles is $\Gamma $-periodic. In this case, the set
of $\Gamma $-periodic density of particles: 
\begin{equation}
S_{per}^{N}=\left\{ n\in S^{N}\text{, }n\left( \vec{x}+\vec{R}\right)
=n\left( \vec{x}\right) \text{ }a.e.\text{, }\vec{R}\in \Gamma \right\}
\end{equation}
is invariant for the map $T$. In this paper, we will search for the fixed
points of the map $T$ only in this invariant set. Thus, from now on, we will
restrict $T$ to $S_{per}^{N}$. For finite volume, the above system include
also the case of nonperiodic background potentials. To include such systems,
the unit cell is taken equal to the entire. Let us denote by $N_{0}=N/K^{3}$
the number of particles per unit cell. The thermodynamic limit is defined by
fixing $N_{0}$ and letting the number of unit cells to go to infinity.

Let us consider the following unitary transformation: 
\begin{gather}
U:L^{2}\left( \mathcal{T}\right) \rightarrow \bigoplus_{\mathbf{q}\in
\Lambda _{K}}L^{2}\left[ cell\right]  \label{Utransformations} \\
L^{2}\left( \mathcal{T}\right) \ni f\rightarrow \bigoplus_{\mathbf{q}\in
\Lambda _{K}}\left( Uf\right) _{\mathbf{q}}  \notag \\
\left( Uf\right) _{\mathbf{q}}\left( \vec{x}\right) =K^{-3/2}\sum_{\mathbf{m}%
\in \Lambda _{K}}e^{-i\Sigma _{j=1}^{3}m^{j}\theta _{\mathbf{q}}^{i}}f\left( 
\vec{x}+\Sigma _{j=1}^{3}m^{j}\vec{\delta}_{j}\right) \text{,}  \notag
\end{gather}
where $\vec{\theta}_{\mathbf{q}}=\dfrac{2\pi }{K}\mathbf{q}\ $and $\Lambda
_{K}=\left\{ 0,1,...,K-1\right\} ^{3}$. For $n\in S_{per}^{N}$, the
Kohn-Sham Hamiltonian is $\Gamma $-periodic and consequently:\cite{SimonIV} 
\begin{equation}
UH_{n}U^{-1}=\bigoplus_{\mathbf{q}\in \Lambda _{K}}\left( -\tfrac{1}{2}%
\Delta _{\vec{\theta}_{\mathbf{q}}}+\lambda V[n]\right) \equiv \bigoplus_{%
\mathbf{q}\in \Lambda _{K}}H_{n}^{\left( \mathbf{q}\right) }\text{,}
\label{Decomposition}
\end{equation}
where $\Delta _{\vec{\theta}}$ is the Laplace operator over the unit cell
with the following boundary conditions: 
\begin{equation}
f\left( \vec{x}+\vec{\delta}_{j}\right) =e^{i\theta ^{j}}f\left( \vec{x}%
\right) \text{ \ and \ }f^{\prime }\left( \vec{x}+\vec{\delta}_{j}\right)
=e^{i\theta ^{j}}f^{\prime }\left( \vec{x}\right)
\end{equation}
for $\vec{x}$ and $\vec{x}+\vec{\delta}_{j}$ on the faces of the unit cell.
The symbol $f^{\prime }$ stands for the derivative of $f$ along $\vec{\delta}%
_{j}$. $V\left[ n\right] $ in Eq. (\ref{Decomposition}) is just the
restriction of the effective potential to the unit cell. Because the kinetic
energy and the effective potential depends on the volume, we will write the
Kohn-Sham Hamiltonian as: 
\begin{equation}
H_{n}^{\left( K\right) }=-\tfrac{1}{2}\Delta +V^{\left( K\right) }\left[ n%
\right] \text{.}
\end{equation}
We will write $T_{K}$ to indicate that the map defined in the previous
section depends on $K$. Also, because $N=K^{3}N_{0}$, it will be more
convenient to use the notation $S_{per}^{K}$ instead of $S_{per}^{N}$.

Let us discuss now the neutrality condition. This condition seems artificial
for interactions other than the Coulomb force. However, when the volume is
transformed to a torus and the particles interact via a potential which
depends only on the distance between the particles, one can immediately see
that adding a uniform background charge has the effect of a constant added
to the Kohn-Sham Hamiltonian or to the full many-body Hamiltonian. This does
not affect the solutions of the Kohn-Sham equations or the physiscs of the
problem. Thus, we can allways add a uniform background charge such that the
neutrality condition Eq. (\ref{Neutrality}) is satisfied. The neutrality
condition can be regarded as a mathematical artifact.

\section{The result}

Our abstract conditions for existence, uniqueness and thermodynamic limit
for the Kohn-Sham equations consist of the following.

\begin{itemize}
\item[(\textit{C}1)]  For any $\vec{\theta}\in \left[ 0,2\pi \right) ^{3}$, $%
n\in S_{per}^{K}$ and $a>0$, 
\begin{equation}
\left\| V^{\left( K\right) }\left[ n\right] \left( -\tfrac{1}{2}\Delta _{%
\vec{\theta}}+a\right) ^{-1}\right\| \leqslant \gamma _{a}\text{,}
\end{equation}
where it is assumed that $\gamma _{a}$ depends only on $a$ (when $N_{0}$ is
kept fixed).

\item[(\textit{C}2)]  There exists a closed set $B\subset L^{1}\left(
cell\right) $ such that $T_{K}\left[ S_{per}^{K}\right] \subset B_{K}$,
where 
\begin{equation}
B_{K}=\left\{ n\in S_{per}^{K}\text{, }\left. n\right| _{cell}\in B\right\} 
\text{,}
\end{equation}
and, for $n_{1,2}\in B_{K}$, there exists a constant $L$, independent of $K$%
, such that: 
\begin{equation}
\left\| V^{\left( K\right) }\left[ n_{1}\right] -V^{\left( K\right) }\left[
n_{2}\right] \right\| _{L^{1}\left( cell\right) }\leqslant L\left\|
n_{1}-n_{2}\right\| _{L^{1}\left( cell\right) }.
\end{equation}

\item[(\textit{C}3)]  For $n\in S_{per}^{K}$, 
\begin{equation}
\left\| V^{\left( K+1\right) }\left[ n\right] -V^{\left( K\right) }\left[ n%
\right] \right\| _{L^{1}\left( cell\right) }\rightarrow 0\text{ \ as \ }%
K\rightarrow \infty \text{,}
\end{equation}
where $n$ in $V^{\left( K+1\right) }\left[ n\right] $ represents the unique
extension of $n$ in $S_{per}^{K+1}$.
\end{itemize}

Our main result is given below.

\begin{theorem}
Suppose that (\textit{C}1)-(\textit{C}3) are satisfied. Then:\newline
\textit{i}) The maps $T_{K}$ are well defined.\newline
\textit{ii}) $T_{K}$ have a unique fixed point provided the coupling
constant is smaller than a certain value which is independent of $K$.\newline
\textit{iii}) The thermodynamic limit of the fixed points is well defined.
\end{theorem}

Because the proof of the first two points and the proof of the third point
of the above Theorem are based on different techniques, we present them in
two separate subsections.

\subsection{Existence and uniqueness for finite volume}

\textit{Proof of Theorem 3. i) }Using the unitary transformation Eq. (\ref
{Utransformations}), for $f\in L^{2}\left( \mathcal{T}\right) $: 
\begin{eqnarray}
&&\left\langle f,V^{\left( K\right) }\left[ n\right] \left( -\tfrac{1}{2}%
\Delta +a\right) ^{-1}f\right\rangle \\
&=&\sum_{\mathbf{q}\in \Lambda _{K}}\left\langle f_{\mathbf{q}},V^{\left(
K\right) }\left[ n\right] \left( -\tfrac{1}{2}\Delta _{\vec{\theta}_{\mathbf{%
q}}}+a\right) ^{-1}f_{\mathbf{q}}\right\rangle  \notag \\
&\leqslant &\gamma _{a}\sum_{\mathbf{q}\in \Lambda _{K}}\left\| f_{\mathbf{q}%
}\right\| _{L^{2}\left( cell\right) }^{2}=\gamma _{a}\left\| f\right\|
_{L^{2}\left( \mathcal{T}\right) }^{2}\text{.}  \notag
\end{eqnarray}
Thus, the conditions of Theorem 1 are uniformly satisfied and in
consequence, all the maps $T_{K}$ are well defined. From this uniform
estimate one can easily conclude that the spectrum of $H_{n}^{(K,\mathbf{q}%
)} $ is uniformly bounded from below, i.e. there exists an $\epsilon _{0}$,
independent of $K$, such that $\sigma \left( H_{n}^{\left( K,\mathbf{q}%
\right) }\right) \subset \left[ \epsilon _{0},\infty \right) $ and,
consequently, $\sigma \left( H_{n}^{\left( K\right) }\right) \subset \left[
\epsilon _{0},\infty \right) $ for all $K$. We can also show that the upper
and lower limits of the chemical potential do not depend on the volume.
Indeed, using again the unitary transformation Eq. (\ref{Utransformations}),
the inequalities Eq. (\ref{miulimits}) can be transformed into: 
\begin{eqnarray}
&&\inf \,Tr\,\phi _{1,\mu _{n}}\left( -\tfrac{1}{2}\left( 1+\lambda \gamma
_{a}\right) \Delta _{\vec{\theta}}+\lambda a\gamma _{a}\right) \\
&\leqslant &N_{0}\leqslant e^{\beta \left( \mu _{n}+\lambda a\gamma
_{a}\right) }\sup \,Tr\,e^{\beta /2\left( 1-\lambda \gamma _{a}\right)
\Delta _{\vec{\theta}}}\text{,}  \notag
\end{eqnarray}
where the infimum and supremum goes over all $\vec{\theta}\in \left[ 0,2\pi
\right) ^{3}$. Using the explicit expressions for the eigenvalues of $\Delta
_{\vec{\theta}}$ and the fact that we can chose $\phi _{1,\mu }$ such that $%
\lim_{\mu \rightarrow \infty }\phi _{1,\mu }=1$, we can conclude from above
that there exist $\mu _{m}$ and $\mu _{M}$, independent of $K$, such that $%
\mu _{n}\in \lbrack \mu _{m},\mu _{M}]$. For $\lambda \leqslant 1$, all
these parameters, $\epsilon _{0}$, $\mu _{m}$ and $\mu _{M}$, can be
considered $\lambda $ independent.$\blacksquare $

We prepare now for the proof of the second point of Theorem 3. Let us prove
an estimate which will be used many times in the following. Along this
paper, $\left\| \cdot \right\| _{1}$ will denote the trace norm.

\begin{proposition}
Let $F$ be an analytic function in a vicinity of $\left[ \epsilon
_{0},\infty \right) $ and $V_{1}$ and $V_{2}$ two self-adjoint potentials
over the unit cell such that: 
\begin{equation}
\left\| V_{1,2}\left( -\tfrac{1}{2}\Delta _{\vec{\theta}}+a\right)
^{-1}\right\| \leqslant \gamma _{a}\text{.}
\end{equation}
Then, 
\begin{eqnarray}
&&\left\| F\left( -\tfrac{1}{2}\Delta _{\vec{\theta}}+\lambda V_{1}\right)
-F\left( -\tfrac{1}{2}\Delta _{\vec{\theta}}+\lambda V_{2}\right) \right\|
_{1} \\
&\leqslant &\lambda \gamma _{F}\left\| V_{1}-V_{2}\right\| _{L^{1}\left(
cell\right) },  \notag
\end{eqnarray}
where $\gamma _{F}$ depends only on the function $F$.
\end{proposition}

\textit{Proof}. Using the notation $\Delta V=V_{1}-V_{2}$ and $g_{a,z}\left(
x\right) =\left( x+a\right) /\left( x-z\right) $, we can write after simple
manipulations: 
\begin{eqnarray}
&&F\left( -\tfrac{1}{2}\Delta _{\vec{\theta}}+\lambda V_{1}\right) -F\left( -%
\tfrac{1}{2}\Delta _{\vec{\theta}}+\lambda V_{2}\right) =\dfrac{\lambda }{%
2\pi i} \\
&&\times \int dz\,F\left( z\right) g_{a,z}\left( -\tfrac{1}{2}\Delta _{\vec{%
\theta}}+\lambda V_{1}\right) \left( 1+\lambda \left( -\tfrac{1}{2}\Delta _{%
\vec{\theta}}+a\right) ^{-1}V_{1}\right) ^{-1}  \notag \\
&&\times \left( -\tfrac{1}{2}\Delta _{\vec{\theta}}+a\right) ^{-1}\Delta
V\left( -\tfrac{1}{2}\Delta _{\vec{\theta}}+a\right) ^{-1}  \notag \\
&&\times \left( 1+\lambda V_{2}\left( -\tfrac{1}{2}\Delta _{\vec{\theta}%
}+a\right) ^{-1}\right) ^{-1}g_{a,z}\left( -\tfrac{1}{2}\Delta _{\vec{\theta}%
}+\lambda V_{2}\right) \text{,}  \notag
\end{eqnarray}
where the integral is on a curve that surrounds $\left[ \epsilon _{0},\infty
\right) $ and belongs to the analyticity domain of $F$. In consequence: 
\begin{eqnarray}
&&\left\| F\left( -\tfrac{1}{2}\Delta _{\vec{\theta}}+V_{1}\right) -F\left( -%
\tfrac{1}{2}\Delta _{\vec{\theta}}+V_{2}\right) \right\| _{1}
\label{Difference} \\
&\leqslant &\dfrac{\lambda }{\left( 1-\lambda \gamma _{a}\right) ^{2}}\int
\left| dz\right| \,\left| F\left( z\right) \right| \sup_{x\in \left[
\epsilon _{0},\infty \right) }\left| g_{a,z}\left( x\right) \right| ^{2} 
\notag \\
&&\times \left\| \left( -\tfrac{1}{2}\Delta _{\vec{\theta}}+a\right)
^{-1}\Delta V\left( -\tfrac{1}{2}\Delta _{\vec{\theta}}+a\right)
^{-1}\right\| _{1}\text{.}  \notag
\end{eqnarray}
Let us consider: 
\begin{equation}
A\equiv \left| \Delta V\right| ^{1/2}\left( -\tfrac{1}{2}\Delta _{\vec{\theta%
}}+a\right) ^{-1}\text{,}
\end{equation}
where the square root is defined through the polar decomposition: $\Delta
V=S\left| \Delta V\right| $. We can immediately see that $A$ is a
Hilbert-Schmidt operator: 
\begin{equation}
\left\| A^{\dagger }A\right\| _{1}=\int_{cell}d\vec{x}\left| \Delta V\left( 
\vec{x}\right) \left( -\tfrac{1}{2}\Delta _{\vec{\theta}}+a\right)
^{-2}\left( \vec{x},\vec{x}\right) \right| \leqslant k_{a}\left\| \Delta
V\right\| _{L^{1}\left( cell\right) }\text{.}  \label{HS}
\end{equation}
We used the fact that $\left( -\tfrac{1}{2}\Delta _{\vec{\theta}}+a\right)
^{-2}\left( \vec{x},\vec{x}\right) $ can be computed explicitly and:\cite
{ProdanHII} 
\begin{equation}
\left( -\tfrac{1}{2}\Delta _{\vec{\theta}}+a\right) ^{-2}\left( \vec{x},\vec{%
x}\right) \leqslant k_{a}\text{,}
\end{equation}
with $k_{a}$ independent of $\vec{\theta}\in \left[ 0,2\pi \right) ^{3}$.
Then we can continue: 
\begin{eqnarray}
&&\left\| \left( -\tfrac{1}{2}\Delta _{\vec{\theta}}+a\right) ^{-1}\Delta
V\left( -\tfrac{1}{2}\Delta _{\vec{\theta}}+a\right) ^{-1}\right\| _{1} \\
&=&\left\| A^{\dagger }SA\right\| _{1}\leqslant \left\| A^{\dagger }\right\|
_{HS}\left\| SA\right\| _{HS}  \notag \\
&\leqslant &\left\| A\right\| _{HS}^{2}=\left\| A^{\dagger }A\right\| _{1}%
\text{,}  \notag
\end{eqnarray}
and this, together with Eqs. (\ref{Difference}) and (\ref{HS}), proves the
affirmation. We can also identify $\gamma _{F}$: 
\begin{equation}
\gamma _{F}=\dfrac{k_{a}}{\left( 1-\lambda \gamma _{a}\right) ^{2}}\int
\left| dz\right| \,\left| F\left( z\right) \right| \sup_{x\in \left[
\epsilon _{0},\infty \right) }\left| g_{a,z}\left( x\right) \right|
^{2}.\blacksquare  \label{gammaF}
\end{equation}

\textit{Proof of Theorem 3.} \textit{ii}). For $n_{1,2}\in B_{K}$ one has: 
\begin{eqnarray}
&&\left\| T\left[ n_{1}\right] -T\left[ n_{2}\right] \right\| _{L^{1}\left( 
\mathcal{T}\right) } \\
&\leqslant &\left\| \Phi _{FD}\left( H_{n_{1}}^{\left( K\right) }-\mu
_{n_{1}}\right) -\Phi _{FD}\left( H_{n_{1}}^{\left( K\right) }-\mu
_{n_{2}}\right) \right\| _{1}  \notag \\
&&+\left\| \Phi _{FD}\left( H_{n_{1}}^{\left( K\right) }-\mu _{n_{2}}\right)
-\Phi _{FD}\left( H_{n_{2}}^{\left( K\right) }-\mu _{n_{2}}\right) \right\|
_{1}  \notag
\end{eqnarray}
Using the monotonicity of the Fermi-Dirac statistics with respect to the
chemical potential we can write: 
\begin{eqnarray}
&&\left\| \Phi _{FD}\left( H_{n_{1}}^{\left( K\right) }-\mu _{n_{1}}\right)
-\Phi _{FD}\left( H_{n_{1}}^{\left( K\right) }-\mu _{n_{2}}\right) \right\|
_{1} \\
&=&\left| Tr\,\Phi _{FD}\left( H_{n_{1}}^{\left( K\right) }-\mu
_{n_{1}}\right) -Tr\,\Phi _{FD}\left( H_{n_{1}}^{\left( K\right) }-\mu
_{n_{2}}\right) \right|  \notag
\end{eqnarray}
At this point we use the fact that: 
\begin{equation}
N=Tr\,\Phi _{FD}\left( H_{n_{1}}^{\left( K\right) }-\mu _{n_{1}}\right)
=Tr\,\Phi _{FD}\left( H_{n_{2}}^{\left( K\right) }-\mu _{n_{2}}\right) \text{%
,}
\end{equation}
so we can conclude: 
\begin{eqnarray}
&&\left\| T\left[ n_{1}\right] -T\left[ n_{2}\right] \right\| _{L^{1}\left( 
\mathcal{T}\right) } \\
&\leqslant &2\left\| \Phi _{FD}\left( H_{n_{1}}^{\left( K\right) }-\mu
_{n_{2}}\right) -\Phi _{FD}\left( H_{n_{2}}^{\left( K\right) }-\mu
_{n_{2}}\right) \right\| _{1}  \notag \\
&=&2\sum_{\mathbf{q}\in \Lambda _{K}}\left\| \Phi _{FD}\left(
H_{n_{1}}^{\left( K,\mathbf{q}\right) }-\mu _{n_{2}}\right) -\Phi
_{FD}\left( H_{n_{2}}^{\left( K,\mathbf{q}\right) }-\mu _{n_{2}}\right)
\right\| _{1}  \notag \\
&\leqslant &2\lambda K^{3}\gamma _{\Phi }\left\| V^{\left( K\right) }\left[
n_{1}\right] -V^{\left( K\right) }\left[ n_{2}\right] \right\| _{L^{1}\left(
cell\right) }  \notag \\
&\leqslant &2\lambda \gamma _{\Phi }L\left\| n_{1}-n_{2}\right\|
_{L^{1}\left( \mathcal{T}\right) }\text{.}  \notag
\end{eqnarray}
A simple analysis of the expression (\ref{gammaF}) shows that $\gamma _{\Phi
}$ is maximum when $\mu _{n_{2}}=\mu _{M}$. $\gamma _{\Phi }$ can be also
chosen $\lambda $ independent for $\lambda \leqslant 1$. Thus we proved
that, for $\lambda $ smaller than a certain constant, independent of $K$,
the maps $T_{K}$ are contractions on the closed, invariant sets $B_{K}$.
This implies that they have a unique fixed point in $B_{K}$. Because $T_{K}%
\left[ S_{per}^{K}\right] \subset B_{K}$, it follows that $T_{K}$ have a
unique fixed point over the entire $S_{per}^{K}$.$\blacksquare $

We end this section with estimates on the chemical potential.

\begin{proposition}
For $\mu _{1,2}\in \left[ \mu _{m},\mu _{M}\right] $ and $n\in S_{per}^{K}$,
there exists $C$ and $C^{\prime }$ strictly positive constants, independent
of $K$ such that: 
\begin{equation}
C\left| \mu _{1}-\mu _{2}\right| \leqslant K^{-3}\left| F_{K}\left( n,\mu
_{1}\right) -F_{K}\left( n,\mu _{2}\right) \right| \leqslant C^{\prime
}\left| \mu _{1}-\mu _{2}\right| \text{,}
\end{equation}
where: 
\begin{equation}
F_{K}\left( n,\mu \right) =Tr\,\Phi _{FD}\left( H_{n}^{\left( K\right) }-\mu
\right) \text{.}
\end{equation}
\end{proposition}

\textit{Proof}. The affirmation follows from: 
\begin{equation}
\left| F_{K}\left( n,\mu _{1}\right) -F_{K}\left( n,\mu _{2}\right) \right|
=\left| \int_{\mu _{1}}^{\mu _{2}}d\mu \dfrac{\partial F_{K}\left( n,\mu
\right) }{\partial \mu }\right|
\end{equation}
and from estimates on 
\begin{equation}
\dfrac{\partial F_{K}\left( n,\mu \right) }{\partial \mu }=\beta e^{-\beta
\mu }Tr\left( 1+e^{\beta (H_{n}^{\left( K\right) }-\mu )}\right) ^{-2}\text{.%
}
\end{equation}
These estimates follows from: 
\begin{eqnarray}
&&\beta e^{-\beta \mu _{M}}Tr\left( 1+e^{\beta (H_{n}^{\left( K\right) }-\mu
_{m})}\right) ^{-2} \\
&\leqslant &\dfrac{\partial F_{K}\left( n,\mu \right) }{\partial \mu }%
\leqslant \beta e^{-\beta \mu _{m}}Tr\left( 1+e^{\beta (H_{n}^{\left(
K\right) }-\mu _{M})}\right) ^{-2}\text{,}  \notag
\end{eqnarray}
which can be reduced to: 
\begin{eqnarray}
&&\beta e^{-\beta \mu _{M}}\inf \,Tr\,\phi _{2,\mu _{m}}\left( -\tfrac{1}{2}%
\left( 1+\lambda \gamma _{a}\right) \Delta _{\vec{\theta}}+\lambda a\gamma
_{a}\right) \\
&\leqslant &K^{-3}\dfrac{\partial F_{K}\left( n,\mu \right) }{\partial \mu }%
\leqslant \beta e^{-\beta \mu _{m}}e^{2\beta \left( \mu _{M}+\lambda a\gamma
_{a}\right) }\sup \,Tr\,e^{\beta \left( 1-\lambda \gamma _{a}\right) \Delta
_{\vec{\theta}}}\text{,}  \notag
\end{eqnarray}
by using the unitary transformation (\ref{Utransformations}) and Lemma 2.
The infimum and supremum go over all $\vec{\theta}\in \left[ 0,2\pi \right)
^{3}$. We can conclude that: 
\begin{equation}
C<K^{-3}\dfrac{\partial F_{K}\left( n,\mu \right) }{\partial \mu }<C^{\prime
}
\end{equation}
where $C$ is strictly positive if $\phi _{2,\mu }>0$, $C^{\prime }<\infty $
and both constants do not depend on $K$. For $\lambda \leqslant 1$, these
constants can be chosen independently of $\lambda $.$\blacksquare $

\begin{theorem}
For any $n\in B_{K}$, the sequence $\left\{ \mu _{T^{\circ m}\left[ n\right]
}\right\} _{m}$ converges to a unique limit.
\end{theorem}

\textit{Proof}. For $n_{1,2}\in B_{K}$, we have successively: 
\begin{eqnarray}
&&CK^{3}\left| \mu _{n_{1}}-\mu _{n_{2}}\right| \\
&\leqslant &\left| F_{K}\left( n_{1},\mu _{n_{1}}\right) -F_{K}\left(
n_{1},\mu _{n_{2}}\right) \right|  \notag \\
&=&\left| N-F_{K}\left( n_{1},\mu _{n_{2}}\right) \right|  \notag \\
&=&\left| F_{K}\left( n_{2},\mu _{n_{2}}\right) -F_{K}\left( n_{1},\mu
_{n_{2}}\right) \right|  \notag \\
&\leqslant &\left\| \Phi _{FD}\left( H_{n_{2}}^{\left( K\right) }-\mu
_{n_{2}}\right) -\Phi _{FD}\left( H_{n_{1}}^{\left( K\right) }-\mu
_{n_{2}}\right) \right\| _{1}  \notag \\
&\leqslant &\lambda K^{3}\gamma _{\Phi }L\left\| n_{1}-n_{2}\right\|
_{L^{1}\left( cell\right) }\text{.}  \notag
\end{eqnarray}
The affirmation follows from the fact that $\left\{ T^{\circ m}\left[ n%
\right] \right\} _{m}$ converges to the same limit for any $n\in S_{per}^{K}$%
.$\blacksquare $

\subsection{The thermodynamic limit}

For a given $K$, let us denote by $n_{K}$ and $\mu _{K}$ the fixed point and
the corresponding chemical potential of the map $T_{K}$. We prepare now to
prove the last point of Theorem 3. We will use the following result from
Ref. \cite{ProdanHII}.

\begin{proposition}
Let $\mu \in \left[ \mu _{m},\mu _{M}\right] $ and $V$ be a self-adjoint
potential such that: 
\begin{equation}
\left\| V\left( -\tfrac{1}{2}\Delta _{\vec{\theta}}+a\right) ^{-1}\right\|
\leqslant \gamma _{a}\text{,}
\end{equation}
for all $\vec{\theta}\in \left[ 0,2\pi \right) ^{3}$. Then: 
\begin{equation}
\left\| \Phi _{FD}\left( -\tfrac{1}{2}\Delta _{\vec{\theta}}+V-\mu \right)
-\Phi _{FD}\left( -\tfrac{1}{2}\Delta _{\vec{\theta}^{\prime }}+V-\mu
\right) \right\| _{1}\leqslant ct.\left| \vec{\theta}-\vec{\theta}^{\prime
}\right| ^{\epsilon }\text{,}
\end{equation}
where $ct.$ and $\epsilon $ depend only on $\gamma _{a}$.
\end{proposition}

The thermodynamic limit of the Kohn-Sam equations will follow from the
following result.

\begin{lemma}
With the above notations, 
\begin{equation}
\left\| n_{K+1}-n_{K}\right\| _{L^{1}\left( cell\right) }\rightarrow 0\text{
\ as \ }K\rightarrow \infty \text{.}
\end{equation}
\end{lemma}

\textit{Proof}. There is a unique extension of $n_{K}$ in $S_{per}^{K+1}$
which will be denoted by the same symbol $n_{K}$. An important observation
is that $n_{K}\in B_{K+1}$. Let us prove first that 
\begin{equation}
\left\| \left[ \Phi _{FD}\left( H_{n_{K}}^{\left( K+1\right) }-\mu
_{K}\right) -\Phi _{FD}\left( H_{n_{K}}^{\left( K\right) }-\mu _{K}\right) %
\right] \left( \vec{x},\vec{x}\right) \right\| _{L^{1}\left( cell\right) }
\label{Preliminary}
\end{equation}
goes to zero as $K$ goes to infinity. Indeed, Eq. (\ref{Preliminary}) can be
evaluated as it follows: 
\begin{align}
& \left\| \tfrac{1}{\left( K+1\right) ^{3}}\sum_{\mathbf{q}\in \Lambda
_{K+1}}\Phi _{FD}\left( H_{n_{K}}^{\left( K+1,\mathbf{q}\right) }-\mu
_{K}\right) \left( \vec{x},\vec{x}\right) -\right. \\
& \left. -\tfrac{1}{K^{3}}\sum_{\mathbf{q}\in \Lambda _{K}}\Phi _{FD}\left(
H_{n_{K}}^{\left( K,\mathbf{q}\right) }-\mu _{K}\right) \left( \vec{x},\vec{x%
}\right) \right\| _{L^{1}\left( cell\right) }  \notag \\
& \leqslant \left[ \left( \tfrac{K+1}{K}\right) ^{3}-1\right] N_{0}+\tfrac{1%
}{\left( K+1\right) ^{3}}\sum_{\mathbf{q}\in \partial \Lambda _{K+1}}\left\|
\Phi _{FD}\left( H_{n_{K}}^{\left( K+1,\mathbf{q}\right) }-\mu _{K}\right)
\right\| _{1}  \notag \\
& +\tfrac{1}{\left( K+1\right) ^{3}}\sum_{\mathbf{q}\in \Lambda _{K}}\left\|
\Phi _{FD}\left( H_{n_{K}}^{\left( K+1,\mathbf{q}\right) }-\mu _{K}\right)
-\Phi _{FD}\left( H_{n_{K}}^{\left( K,\mathbf{q}\right) }-\mu _{K}\right)
\right\| _{1}\text{,}  \notag
\end{align}
where $\partial \Lambda _{K+1}=\Lambda _{K+1}\backslash \Lambda _{K}$. The
first two terms above go to zero as $K$ goes to infinity. For the last term
we use the following: 
\begin{eqnarray}
&&\left\| \Phi _{FD}\left( -\tfrac{1}{2}\Delta _{\vec{\theta}}+V-\mu \right)
-\Phi _{FD}\left( -\tfrac{1}{2}\Delta _{\vec{\theta}^{\prime }}+V^{\prime
}-\mu \right) \right\| _{1} \\
&\leqslant &\left\| \Phi _{FD}\left( -\tfrac{1}{2}\Delta _{\vec{\theta}%
}+V-\mu \right) -\Phi _{FD}\left( -\tfrac{1}{2}\Delta _{\vec{\theta}%
}+V^{\prime }-\mu \right) \right\| _{1}  \notag \\
&&+\left\| \Phi _{FD}\left( -\tfrac{1}{2}\Delta _{\vec{\theta}}+V^{\prime
}-\mu _{K}\right) -\Phi _{FD}\left( -\tfrac{1}{2}\Delta _{\vec{\theta}%
^{\prime }}+V^{\prime }-\mu \right) \right\| _{1}  \notag \\
&\leqslant &\lambda \gamma _{\Phi }\left\| V-V^{\prime }\right\|
_{L^{1}\left( cell\right) }+ct.\left| \vec{\theta}-\vec{\theta}^{\prime
}\right| ^{\epsilon }\text{.}  \notag
\end{eqnarray}
Replacing $\vec{\theta}$ and $\vec{\theta}^{\prime }$ by $2\pi \mathbf{q}%
/\left( K+1\right) $ and $2\pi \mathbf{q}/K$, $V$ and $V^{\prime }$ by $%
V^{\left( K+1\right) }\left[ n_{K}\right] $ and $V^{\left( K\right) }\left[
n_{K}\right] $ and $\mu $ by $\mu _{K}$, Eq. (\ref{Preliminary}) follows
immediately from condition (\textit{C}3).

Let us denote by $\mu _{n_{K}}$ be the chemical potential corresponding to $%
n_{K}$ when $n_{K}$ is extended in $B_{K+1}$, i.e.: 
\begin{equation}
\left( K+1\right) ^{3}N_{0}=Tr\,\Phi _{FD}\left( H_{n_{K}}^{\left(
K+1\right) }-\mu _{n_{K}}\right) \text{.}
\end{equation}
It follows that: 
\begin{equation}
\left| \mu _{K}-\mu _{n_{K}}\right| \rightarrow 0\text{ \ as }K\rightarrow
\infty \text{.}  \label{miulimit}
\end{equation}
Indeed, from Proposition 5, 
\begin{eqnarray}
&&C\left| \mu _{K}-\mu _{n_{K}}\right| \leqslant \tfrac{1}{\left( K+1\right)
^{3}} \\
&&\times \left| Tr\,\Phi _{FD}\left( H_{n_{K}}^{\left( K+1\right) }-\mu
_{K}\right) -Tr\,\Phi _{FD}\left( H_{n_{K}}^{\left( K+1\right) }-\mu
_{n_{K}}\right) \right|  \notag \\
&\leqslant &\left| \tfrac{1}{\left( K+1\right) ^{3}}Tr\,\Phi _{FD}\left(
H_{n_{K}}^{\left( K+1\right) }-\mu _{K}\right) -N_{0}\right|  \notag \\
&\leqslant &\left| \tfrac{1}{\left( K+1\right) ^{3}}Tr\,\Phi _{FD}\left(
H_{n_{K}}^{\left( K+1\right) }-\mu _{K}\right) -\tfrac{1}{K^{3}}Tr\,\Phi
_{FD}\left( H_{n_{K}}^{\left( K\right) }-\mu _{K}\right) \right|  \notag \\
&\leqslant &\left\| \left[ \Phi _{FD}\left( H_{n_{K}}^{\left( K+1\right)
}-\mu _{K}\right) -\Phi _{FD}\left( H_{n_{K}}^{\left( K\right) }-\mu
_{K}\right) \right] \left( \vec{x},\vec{x}\right) \right\| _{L^{1}\left(
cell\right) }\text{.}  \notag
\end{eqnarray}
Then Eq. (\ref{miulimit}) follows from Eq. (\ref{Preliminary}).

Finally, 
\begin{eqnarray}
&&\left\| n_{K+1}-n_{K}\right\| _{L^{1}\left( cell\right) } \\
&=&\left\| \sum_{m=1}^{\infty }\left( T_{K+1}^{\circ m}\left[ n_{K}\right]
-T_{K+1}^{\circ (m-1)}\left[ n_{K}\right] \right) \right\| _{L^{1}\left(
cell\right) }  \notag \\
&\leqslant &\left( 1-2\lambda L\gamma _{\Phi }\right) ^{-1}\left\| T_{K+1} 
\left[ n_{K}\right] -n_{K}\right\| _{L^{1}\left( cell\right) }  \notag
\end{eqnarray}
and: 
\begin{eqnarray}
&&\left\| T_{K+1}\left[ n_{K}\right] -n_{K}\right\| _{L^{1}\left(
cell\right) }  \label{Finally} \\
&=&\left\| \left[ \Phi _{FD}\left( H_{n_{K}}^{\left( K+1\right) }-\mu
_{n_{K}}\right) -\Phi _{FD}\left( H_{n_{K}}^{\left( K\right) }-\mu
_{K}\right) \right] \left( \vec{x},\vec{x}\right) \right\| _{L^{1}\left(
cell\right) }  \notag \\
&\leqslant &\left\| \left[ \Phi _{FD}\left( H_{n_{K}}^{\left( K+1\right)
}-\mu _{n_{K}}\right) -\Phi _{FD}\left( H_{n_{K}}^{\left( K+1\right) }-\mu
_{K}\right) \right] \left( \vec{x},\vec{x}\right) \right\| _{L^{1}\left(
cell\right) }  \notag \\
&&+\left\| \left[ \Phi _{FD}\left( H_{n_{K}}^{\left( K+1\right) }-\mu
_{K}\right) -\Phi _{FD}\left( H_{n_{K}}^{\left( K\right) }-\mu _{K}\right) %
\right] \left( \vec{x},\vec{x}\right) \right\| _{L^{1}\left( cell\right) }%
\text{.}  \notag
\end{eqnarray}
We already proved that the last term goes to zero as $K\rightarrow \infty $.
Using the monotonicity of the Fermi-Dirac, the second last term of Eq. (\ref
{Finally}) is equal to 
\begin{equation}
\left( K+1\right) ^{-3}\left| F_{K+1}\left( n_{K},\mu _{n_{K}}\right)
-F_{K+1}\left( n_{K},\mu _{K}\right) \right| \leqslant C^{\prime }\left| \mu
_{n_{K}}-\mu _{K}\right| \text{,}
\end{equation}
where we used the same notations as in Proposition 5.$\blacksquare $

Within our conditions, we cannot prove that $\left\{ n_{K}\right\} _{K}$ is
a Cauchy sequence and in consequence the problem of the thermodynamic limit
is not yet solved. However, using the weaker result of the above Lemma, we
can prove that the density of particle converges in a distributional sense
as the thermodynamic limit is considered. This will end the proof of Theorem
3.

\begin{theorem}
If $n_{K}$ is viewed as a linear functional over $L^{\infty }\left(
cell\right) $, 
\begin{equation}
\hat{n}_{K}\left( g\right) =\int_{cell}n_{K}\left( \vec{x}\right) g\left( 
\vec{x}\right) d\vec{x}\text{ , \ }g\in L^{\infty }\left( cell\right) \text{,%
}
\end{equation}
then $\left\{ \hat{n}_{K}\right\} _{K}$ converges weakly in $L^{\infty
}\left( cell\right) ^{\ast }$.
\end{theorem}

\textit{Proof}. From Banach-Alaoglu theorem \cite{SimonI} one knows that the
closed balls in $L^{\infty }\left( cell\right) ^{\ast }$ are compact in the
weak topology. Then, because $\left\| \hat{n}_{K}\right\| =N_{0}$, it
follows that the sequence $\left\{ \hat{n}_{K}\right\} _{K}$ has at least
one accumulation point. Due to the fact that 
\begin{equation}
\left\| \hat{n}_{K+1}-\hat{n}_{K}\right\| =\left\| n_{K+1}-n_{K}\right\|
_{L^{1}\left( cell\right) }\underset{K\rightarrow \infty }{\longrightarrow }0%
\text{,}
\end{equation}
we can conclude that there is one and only one accumulation point.$%
\blacksquare $

\section{Application: The Local Density Approximation}

In the local density approximation, the effective potential becomes: 
\begin{equation}
V\left[ n\right] =\left( n-n_{0}\right) \ast v+v_{xc}\left( n\right) \text{,}
\end{equation}
where $v_{xc}$ is a function of $n$ instead of a functional. The value of $%
v_{xc}\left( n\right) $ is equal to the exchange-correlation energy per
particle of the corresponding infinite, homogeneous system. Thus, there will
be no volume dependence for the exchange-correlation potential. The Hartree
potential however will depend, in general, on the volume. For finite range
interactions, the Hartree potential does not depend on the volume when the
volume becomes larger than the range of the interaction. This case has been
considered in Ref. \cite{ProdanHII} when the thermodynamic limit of the
Hartree model was analyzed. We impose the following conditions on the
two-body interaction and exchange-correlation potential.

\begin{itemize}
\item[(\textit{P}1)]  The singularity of the two-body potential is at least $%
L^{2}$ integrable.

\item[(\textit{P}2)]  $v\left( \vec{x}\right) \sim \left| \vec{x}\right|
^{-r}$ as $\left| \vec{x}\right| \rightarrow \infty $ with $r>2$.

\item[(\textit{P}3)]  $v_{xc}:\left( 0,\infty \right) \rightarrow R$ is
differentiable and there exists $p\geqslant 1$ such that $%
t^{-1/p}v_{xc}\left( t\right) $ and $t^{1-1/p}\dfrac{dv_{xc}}{dt}\left(
t\right) $ are uniformly bounded over $[0,\infty )$.
\end{itemize}

The conditions we imposed on the exchange-correlation include, for example,
the case of an homogeneous electron gas. For this particular system, it was
shown \cite{Perdew92} that the low and high density behavior of the
exchange-correlation potential is dominated by the exchange part which is
proportional to $n^{1/3}$. Thus, we can choose $p=3$ in (\textit{P}3) to
include this case. Let us mention that, because of this behavior, the
difference $v_{xc}\left( n_{1}\right) -v_{xc}\left( n_{2}\right) $ decays
much slower than $n_{1}-n_{2}$ in the low density limit. Thus, the condition
(\textit{C}2) fails for this particular potential unless we can prove that
the density of particles is larger than a certain strictly positive value.

\subsection{Estimates on the Hartree potential}

Let us point out that it is the Hartree potential that forces on us to
consider only short range interactions. If one compares the conditions (%
\textit{P}1)-(\textit{P}3) with the conditions from Ref. \cite{ProdanHI},
one can see an improvement because now we allow the interaction to decay as $%
\left| \vec{x}\right| ^{-r}$ with $r>2$ instead of $3$. The neutrality
condition will play an essential role here. Unfortunately, we are still
unable to include the Coulomb interaction in our theory. The reason is that
the Hartree potential increases too fast as the system approaches the
thermodynamic limit for long range interactions.

\begin{theorem}
Suppose the conditions (\textit{P}1) and (\textit{P}2) are satisfied. Then,
for $n$, $n^{\prime }\in S_{per}^{K}$ and $\vec{\theta}\in \lbrack 0,2\pi
)^{3}$ the following are true:\newline
\textit{i}) There exists $\gamma _{a}^{H}$, independent of $K$ or $\vec{%
\theta}\in \lbrack 0,2\pi )^{3}$, such that: 
\begin{equation}
\left\| \left( n-n_{0}\right) \ast v\left( -\tfrac{1}{2}\Delta _{\vec{\theta}%
}+a\right) ^{-1}\right\| \leqslant \gamma _{a}^{H}\text{.}
\end{equation}
\textit{ii}) There exists $L^{H}$, independent of $K$ or $\vec{\theta}\in
\lbrack 0,2\pi )^{3}$, such that: 
\begin{equation}
\left\| \left( n-n^{\prime }\right) \ast v\right\| _{L^{1}\left( cell\right)
}\leqslant L^{H}\left\| n-n^{\prime }\right\| _{L^{1}\left( cell\right) }%
\text{.}
\end{equation}
\textit{iii}) The Hartree potential satisfies the condition (\textit{C}3).
\end{theorem}

\textit{Proof}. Let us divide the Hartree potential in two parts, 
\begin{equation}
\sum_{i=1,2}\int_{vol^{\left( i\right) }}v\left( \left| \vec{x},\vec{y}%
\right| \right) \left( n\left( \vec{y}\right) -n_{0}\left( \vec{y}\right)
\right) d\vec{y}\equiv V_{H}^{\left( 1\right) }\left[ n\right]
+V_{H}^{\left( 2\right) }\left[ n\right] \text{,}
\end{equation}
where $vol^{\left( 1\right) }$ contains the unit cell plus the adjacent
cells and $vol^{\left( 2\right) }=\mathcal{T\,}\backslash \,vol^{\left(
1\right) }$. For $n$, $n^{\prime }\in S_{per}^{K}$, it follows 
\begin{eqnarray}
&&\left\| V_{H}^{\left( 1\right) }\left[ n\right] -V_{H}^{\left( 1\right) }%
\left[ n^{\prime }\right] \right\| _{L^{1}\left( cell\right) } \\
&\leqslant &\mathcal{N}\sup_{\vec{y}\in vol^{\left( 1\right) }}\int_{cell}d%
\vec{x}\,\left| v\left( \left| \vec{x},\vec{y}\right| \right) \right|
\,\left\| n-n^{\prime }\right\| _{L^{1}\left( cell\right) }\text{.}  \notag
\end{eqnarray}
where $\mathcal{N}$ is the number of cells in $vol^{\left( 1\right) }$. The
estimate makes sense because, if the singularity of the interacting
potential is $L^{2}$ integrable, then it is also $L^{1}$ integrable.
Moreover, 
\begin{equation*}
\left\| V_{H}^{\left( 1\right) }\left[ n\right] \right\| _{L^{2}\left(
cell\right) }\leqslant 2\mathcal{N}N_{0}\sqrt{\sup \int_{cell}d\vec{x}%
\,\left| v\left( \vec{y}_{1},\vec{x}\right) v\left( \left| \vec{x},\vec{y}%
_{2}\right| \right) \right| }\text{,} 
\end{equation*}
where the supremum goes over all $\vec{y}_{1}$ and $\vec{y}_{2}\in
vol^{\left( 1\right) }$. Again, the estimate makes sense because the
singularity of the interaction is $L^{2}$ integrable. The last inequality
combined with 
\begin{equation}
\left\| \left( -\tfrac{1}{2}\Delta _{\vec{\theta}}+a\right) ^{-1}f\right\|
_{L^{\infty }\left( cell\right) }\leqslant k_{a}^{1/2}\left\| f\right\|
_{L^{2}\left( cell\right) }  \label{Basic}
\end{equation}
from Ref. \cite{ProdanHII}, leads to: 
\begin{eqnarray}
&&\left\| V_{H}^{\left( 1\right) }\left[ n\right] \left( -\tfrac{1}{2}\Delta
_{\vec{\theta}}+a\right) ^{-1}f\right\| _{L^{2}\left( cell\right) } \\
&\leqslant &\left\| V_{H}^{\left( 1\right) }\left[ n\right] \right\|
_{L^{2}\left( cell\right) }\left\| \left( -\tfrac{1}{2}\Delta _{\vec{\theta}%
}+a\right) ^{-1}f\right\| _{L^{\infty }\left( cell\right) }  \notag \\
&\leqslant &2k_{a}^{1/2}\mathcal{N}N_{0}\sqrt{\sup_{\vec{y}\in vol^{\left(
1\right) }}\int_{cell}d\vec{x}\,\left| v\left( \left| \vec{x},\vec{y}\right|
\right) \right| ^{2}}\left\| f\right\| _{L^{2}\left( cell\right) }\text{.} 
\notag
\end{eqnarray}
For the second term we write 
\begin{eqnarray}
&&V_{H}^{\left( 2\right) }\left[ n\right] \left( \vec{x}\right)
-V_{H}^{\left( 2\right) }\left[ n^{\prime }\right] \left( \vec{x}\right) \\
&=&\sum_{\vec{R}}\int_{cell}v\left( \left| \vec{x},\vec{y}+\vec{R}\right|
\right) \left( n\left( \vec{y}\right) -n^{\prime }\left( \vec{y}\right)
\right) d\vec{y}  \notag \\
&=&\sum_{\vec{R}}\int_{cell}\left[ v\left( \left| \vec{x},\vec{y}+\vec{R}%
\right| \right) -v\left( \left| \vec{x},\vec{R}\right| \right) \right]
\left( n\left( \vec{y}\right) -n^{\prime }\left( \vec{y}\right) \right) d%
\vec{y}\text{,}  \notag
\end{eqnarray}
where the sum goes over all the sites of the crystal less the origin and its
first neighbors. It is this place where we used the neutrality condition.
Then: 
\begin{eqnarray}
&&\left| V_{H}^{\left( 2\right) }\left[ n\right] \left( \vec{x}\right)
-V_{H}^{\left( 2\right) }\left[ n^{\prime }\right] \left( \vec{x}\right)
\right|  \label{HartreeDiff} \\
&\leqslant &\sum_{\vec{R}}\sup_{\vec{x},\vec{y}\in cell}\left| v\left(
\left| \vec{x},\vec{y}+\vec{R}\right| \right) -v\left( \left| \vec{x},\vec{R}%
\right| \right) \right| \left\| n-n^{\prime }\right\| _{L^{1}\left(
cell\right) }  \notag \\
&\leqslant &\sum_{\vec{R}}\sup_{\vec{x},\vec{y}\in cell}\left| v\left( \vec{R%
}-\vec{x}+\vec{y}\right) -v\left( \vec{R}-\vec{x}\right) \right| \left\|
n-n^{\prime }\right\| _{L^{1}\left( cell\right) }\text{,}  \notag
\end{eqnarray}
where the last sum goes over an infinite lattice. Denoting $\vec{\xi}=\vec{R}%
/R$, $R=|\vec{R}|$, it follows from condition (\textit{P}2) that, for large $%
R$, 
\begin{eqnarray}
&&\left| v\left( \vec{R}-\vec{x}+\vec{y}\right) -v\left( \vec{R}-\vec{x}%
\right) \right| \\
&=&ct.R^{-r}\left| \left| \vec{\xi}-\left( \vec{x}-\vec{y}\right) /R\right|
^{-r}-\left| \vec{\xi}-\vec{x}/R\right| ^{-r}\right|  \notag \\
&\leqslant &ct.R^{-r}\left| \left| \vec{\xi}-\left( \vec{x}-\vec{y}\right)
/R\right| -\left| \vec{\xi}-\vec{x}/R\right| \right|  \notag
\end{eqnarray}
and using $\left| |\vec{a}+\vec{b}|-|\vec{a}-\vec{b}|\right| \leqslant 2|%
\vec{b}|$ we can conclude: 
\begin{equation}
\left| v\left( \vec{R}-\vec{x}+\vec{y}\right) -v\left( \vec{R}-\vec{x}%
\right) \right| \leqslant ct.R^{-r-1}\text{.}  \label{Decay}
\end{equation}
Thus the sum in Eq. (\ref{HartreeDiff}) converges and we proved that there
exists $L_{H}^{\left( 2\right) }$, independent of $K$, such that: 
\begin{equation}
\left\| V_{H}^{\left( 2\right) }\left[ n\right] -V_{H}^{\left( 2\right) }%
\left[ n^{\prime }\right] \right\| _{L^{\infty }\left( cell\right)
}\leqslant L_{H}^{\left( 2\right) }\left\| n-n^{\prime }\right\|
_{L^{1}\left( cell\right) }
\end{equation}
which automatically leads to: 
\begin{equation}
\left\| V_{H}^{\left( 2\right) }\left[ n\right] -V_{H}^{\left( 2\right) }%
\left[ n^{\prime }\right] \right\| _{L^{1}\left( cell\right) }\leqslant
v_{cell}L_{H}^{\left( 2\right) }\left\| n-n^{\prime }\right\| _{L^{1}\left(
cell\right) }
\end{equation}
and 
\begin{equation}
\left\| V_{H}^{\left( 2\right) }\left[ n\right] \left( -\tfrac{1}{2}\Delta _{%
\vec{\theta}}+a\right) ^{-1}\right\| \leqslant
2k_{a}^{1/2}N_{0}v_{cell}^{1/2}L_{H}^{\left( 2\right) }\text{.}
\end{equation}
This ends the proof of point \textit{i)} and \textit{ii}). For point \textit{%
iii}), we notice that the difference between $V_{H}^{\left( K\right) }[n]$
and $V_{H}^{\left( K+1\right) }\left[ n\right] $ is given by: 
\begin{equation*}
\sum_{\vec{R}\in \partial \Lambda _{K+1}}\int_{cell}\left[ v\left( \left| 
\vec{x},\vec{y}+\vec{R}\right| \right) -v\left( \left| \vec{x},\vec{R}%
\right| \right) \right] \left( n\left( \vec{y}\right) -n_{0}\left( \vec{y}%
\right) \right) d\vec{y}\text{,} 
\end{equation*}
and the sum contains a number of terms proportional to $\left( K+1\right)
^{2}$ and $R$ is proportional to $K$. Then the affirmation follows from Eq. (%
\ref{Decay}).$\blacksquare $

\subsection{Estimates on the exchange-correlation potential}

Condition (\textit{P}3) automatically leads to the condition (\textit{C}1)
for the exchange-correlation potential.

\begin{proposition}
For $n\in S_{per}^{K}$, 
\begin{equation}
\left\| v_{xc}\left( n\right) \left( -\tfrac{1}{2}\Delta _{\vec{\theta}%
}+a\right) ^{-1}\right\| \leqslant \gamma _{a}^{xc}\text{.}
\end{equation}
\end{proposition}

\textit{Proof}. 
\begin{eqnarray}
&&\left\| v_{xc}\left( n\right) \right\| _{L^{p}\left( cell\right) } \\
&=&\left[ \int_{cell}\left| n\left( \vec{x}\right) ^{-1/p}v_{xc}\left(
n\left( \vec{x}\right) \right) \right| ^{p}n\left( \vec{x}\right) d\vec{x}%
\right] ^{1/p}  \notag \\
&\leqslant &\sup_{t\in R_{+}}t^{-1/p}v_{xc}\left( t\right) \left\| n\right\|
_{L^{1}\left( cell\right) }^{1/p}  \notag \\
&\leqslant &N_{0}^{1/p}\sup_{t\in R_{+}}t^{-1/p}v_{xc}\left( t\right) \text{,%
}  \notag
\end{eqnarray}
and using Eq. (\ref{Basic}) 
\begin{eqnarray}
&&\left\| v_{xc}\left( n\right) \left( -\tfrac{1}{2}\Delta _{\vec{\theta}%
}+a\right) ^{-1}f\right\| _{L^{2}\left( cell\right) } \\
&\leqslant &\left\| v_{xc}\left( n\right) \right\| _{L^{p}\left( cell\right)
}\left\| \left( -\tfrac{1}{2}\Delta _{\vec{\theta}}+a\right) ^{-1}f\right\|
_{L^{q}\left( cell\right) }  \notag \\
&\leqslant &v_{cell}^{1/2-1/p}\left\| v_{xc}\left( n\right) \right\|
_{L^{p}\left( cell\right) }\left\| \left( -\tfrac{1}{2}\Delta _{\vec{\theta}%
}+a\right) ^{-1}f\right\| _{L^{\infty }\left( cell\right) }  \notag \\
&\leqslant &k_{a}^{1/2}v_{cell}^{1/2-1/p}\left\| v_{xc}\left( n\right)
\right\| _{L^{p}\left( cell\right) }\left\| f\right\| _{L^{2}\left(
cell\right) }\text{.}\blacksquare  \notag
\end{eqnarray}
Combining the above result with the results from the previous section, it
follows that the condition (\textit{C}1) is satisfied for the local density
approximation of the effective potential. We can also prove directly that
the map $T$ is continuous on $S_{per}^{K}$.

\begin{proposition}
Let $p\geqslant 1$. Then 
\begin{eqnarray}
&&\left\| v_{xc}\left( n_{1}\right) -v_{xc}\left( n_{2}\right) \right\|
_{L^{p}\left( cell\right) } \\
&\leqslant &p\sup_{t\in R_{+}}\left| t^{1-1/p}v_{xc}^{\prime }\left(
t\right) \right| \left\| n_{1}-n_{2}\right\| _{L^{1}\left( cell\right)
}^{1/p}\text{,}  \notag
\end{eqnarray}
and consequently: 
\begin{eqnarray}
&&\left\| T_{K}\left[ n_{1}\right] -T_{K}\left[ n_{2}\right] \right\|
_{L^{1}\left( \mathcal{T}\right) } \\
&\leqslant &2\lambda \gamma _{\Phi }L^{H}\left\| n_{1}-n_{2}\right\|
_{L^{1}\left( \mathcal{T}\right) }  \notag \\
&&+2\lambda \gamma _{\Phi }pK^{3-3/p}\sup_{t\in R_{+}}\left|
t^{1-1/p}v_{xc}^{\prime }\left( t\right) \right| \left\| n_{1}-n_{2}\right\|
_{L^{1}\left( \mathcal{T}\right) }^{1/p}\text{,}  \notag
\end{eqnarray}
for $n_{1,2}\in S_{per}^{K}$.
\end{proposition}

\textit{Proof}. For $t_{1}$, $t_{2}\in R_{+}$%
\begin{eqnarray*}
v_{xc}\left( t_{1}\right) -v_{xc}\left( t_{2}\right)
&=&\int_{t_{1}^{1/p}}^{t_{2}^{1/p}}\frac{dv_{xc}}{dt^{1/p}}dt^{1/p} \\
&\leqslant &p\sup_{t\in R_{+}}\left| t^{1-1/p}v_{xc}^{\prime }\left(
t\right) \right| \left| t_{1}^{1/p}-t_{2}^{1/p}\right| \\
&\leqslant &p\sup_{t\in R_{+}}\left| t^{1-1/p}v_{xc}^{\prime }\left(
t\right) \right| \left| t_{1}-t_{2}\right| ^{1/p}\text{,}
\end{eqnarray*}
for $p\geqslant 1$. We can continue: 
\begin{eqnarray}
&&\left\| v_{xc}\left( n_{1}\right) -v_{xc}\left( n_{2}\right) \right\|
_{L^{p}\left( cell\right) } \\
&\leqslant &\left( \int_{cell}\left( p\sup_{t\in R_{+}}\left|
t^{1-1/p}v_{xc}^{\prime }\left( t\right) \right| \left| n_{1}\left( \vec{x}%
\right) -n_{2}\left( \vec{x}\right) \right| ^{1/p}\right) ^{p}d\vec{x}%
\right) ^{1/p}  \notag \\
&=&p\sup_{t\in R_{+}}\left| t^{1-1/p}v_{xc}^{\prime }\left( t\right) \right|
\left\| n_{1}-n_{2}\right\| _{L^{1}\left( cell\right) }^{1/p}\text{.}  \notag
\end{eqnarray}
An important consequence of the above result is that any $L^{q}$ norm with $%
q\leqslant p$ of $v_{xc}\left( n_{1}\right) -v_{xc}\left( n_{2}\right) $ is
finite. In particular: 
\begin{eqnarray}
&&\left\| v_{xc}\left( n_{1}\right) -v_{xc}\left( n_{2}\right) \right\|
_{L^{1}\left( cell\right) } \\
&\leqslant &pv_{cell}^{1-1/p}\sup_{t\in R_{+}}\left| t^{1-1/p}v_{xc}^{\prime
}\left( t\right) \right| \left\| n_{1}-n_{2}\right\| _{L^{1}\left(
cell\right) }^{1/p}.  \notag
\end{eqnarray}
We notice that the limits on the chemical potential were based only on the
condition (\textit{C}1). Then, following the steps of the proof of point 
\textit{ii}), Theorem 3, and Proposition 4 we have successively: 
\begin{eqnarray}
&&\left\| T_{K}\left[ n_{1}\right] -T_{K}\left[ n_{2}\right] \right\|
_{L^{1}\left( \mathcal{T}\right) } \\
&\leqslant &2\sum_{\mathbf{q}\in \Lambda _{K}}\left\| \Phi _{FD}\left(
H_{n_{1}}^{\left( K,\mathbf{q}\right) }-\mu _{n_{2}}\right) -\Phi
_{FD}\left( H_{n_{2}}^{\left( K,\mathbf{q}\right) }-\mu _{n_{2}}\right)
\right\| _{1}  \notag \\
&\leqslant &2\lambda \gamma _{\Phi }\left( \left\| \left( n_{1}-n_{2}\right)
\ast v\right\| _{L^{1}\left( \mathcal{T}\right) }+\left\| v_{xc}\left(
n_{1}\right) -v_{xc}\left( n_{2}\right) \right\| _{L^{1}\left( \mathcal{T}%
\right) }\right) \text{.}\blacksquare  \notag
\end{eqnarray}
Of course, this result is far from condition (\textit{C}3) because, for
realistic exchange-correlation potentials we must choose $p\geqslant 1$
above. The above result however, is the best estimate one can get on the map 
$T$ if only $L^{1}$ estimates on the density of particles are used. To
complete our analysis we need $L^{\infty }$ estimates on the density of
particles.

\begin{lemma}
Let $A$ and $W$ be two self-adjoint operators on $L^{2}\left( cell\right) $
such that $A^{-1}$ exists, 
\begin{equation}
\sup_{\vec{x},\vec{y}\in cell}\left| A^{-1}\left( \vec{x},\vec{y}\right)
\right| \leqslant \chi <\infty \text{,}
\end{equation}
and $AWA$ is bounded. Then: 
\begin{equation}
\left| W\left( \vec{x},\vec{x}\right) \right| \leqslant v_{cell}\chi
^{2}\left\| AWA\right\| \text{.}
\end{equation}
\end{lemma}

\textit{Proof}. From the above conditions it follows that 
\begin{equation}
f\left( \vec{y}\right) =A^{-1}\left( \vec{y},\vec{x}\right) \in L^{2}\left(
cell\right) \text{.}
\end{equation}
Thus: 
\begin{eqnarray}
\left| W\left( \vec{x},\vec{x}\right) \right| &=&\left\langle f,\left(
AWA\right) f\right\rangle \leqslant \left\| AWA\right\| \left\| f\right\|
_{L^{2}\left( cell\right) }^{2} \\
&\leqslant &v_{cell}\chi ^{2}\left\| AWA\right\| \text{.}\blacksquare  \notag
\end{eqnarray}

\begin{theorem}
For $n\in S_{per}^{K}$, 
\begin{equation}
n_{\min }-ct.\lambda \leqslant \left\| T_{K}\left[ n\right] \right\|
_{L^{\infty }}\leqslant n_{\max }+ct.\lambda \text{,}
\end{equation}
where: 
\begin{eqnarray}
n_{\min } &=&\dfrac{1}{v_{cell}}\inf_{\vec{\theta}\in \lbrack 0,2\pi
)^{3}}\,Tr\,\Phi _{FD}\left( -\tfrac{1}{2}\Delta _{\vec{\theta}}-\mu
_{m}\right) \\
n_{\max } &=&\dfrac{1}{v_{cell}}\sup_{\vec{\theta}\in \lbrack 0,2\pi
)^{3}}\,Tr\,\Phi _{FD}\left( -\tfrac{1}{2}\Delta _{\vec{\theta}}-\mu
_{M}\right) \text{.}  \notag
\end{eqnarray}
and $ct$. is $K$ independent.
\end{theorem}

\textit{Proof}. In the previous Lemma, we choose: 
\begin{equation}
W_{\mathbf{q}}=\Phi _{FD}\left( H_{n}^{\left( K,\mathbf{q}\right) }-\mu
_{n}\right) -\Phi _{FD}\left( -\tfrac{1}{2}\Delta _{\vec{\theta}_{\mathbf{q}%
}}-\mu _{n}\right)
\end{equation}
and 
\begin{equation}
A_{\mathbf{q}}=\left( -\tfrac{1}{2}\Delta _{\vec{\theta}_{\mathbf{q}%
}}+a\right) ^{2}\text{.}
\end{equation}
In this case $A_{\mathbf{q}}^{-1}$ can be computed exactly\cite{ProdanHII}
and 
\begin{equation}
\left| A_{\mathbf{q}}^{-1}\left( \vec{x},\vec{y}\right) \right| \leqslant 
\frac{1}{2\pi \sqrt{2a}}\sum_{\vec{R}\in \Gamma }e^{-2a\left| \vec{x}-\vec{y}%
-\vec{R}\right| }\text{.}
\end{equation}
Thus, the conditions on $A_{\mathbf{q}}$ in the previous Lemma are
satisfied. Moreover, because 
\begin{gather}
W_{\mathbf{q}}=\tfrac{1}{2}\left( e^{-\frac{\beta }{2}H_{n}^{\left( K,%
\mathbf{q}\right) }}-e^{-\frac{\beta }{2}(-\frac{1}{2}\Delta _{\vec{\theta}_{%
\mathbf{q}}})}\right) \cosh \left[ \tfrac{\beta }{2}\left( H_{n}^{\left( K,%
\mathbf{q}\right) }-\mu _{n}\right) \right] ^{-1} \\
+\tfrac{1}{2}e^{\frac{\beta }{4}\Delta _{\vec{\theta}_{\mathbf{q}}}}\left(
\cosh \left[ \tfrac{\beta }{2}\left( H_{n}^{\left( K,\mathbf{q}\right) }-\mu
_{n}\right) \right] ^{-1}-\cosh \left[ \tfrac{\beta }{2}\left( -\tfrac{1}{2}%
\Delta _{\vec{\theta}_{\mathbf{q}}}-\mu _{n}\right) \right] ^{-1}\right) 
\notag
\end{gather}
and 
\begin{equation}
\cosh \left[ \tfrac{\beta }{2}\left( H_{n}^{\left( K,\mathbf{q}\right) }-\mu
_{n}\right) \right] ^{-1}\left( -\tfrac{1}{2}\Delta _{\vec{\theta}_{\mathbf{q%
}}}+a\right) ^{2}
\end{equation}
and 
\begin{equation}
\left( -\tfrac{1}{2}\Delta _{\vec{\theta}_{\mathbf{q}}}+a\right) ^{2}e^{%
\frac{\beta }{4}\Delta _{\vec{\theta}_{\mathbf{q}}}}
\end{equation}
are of trace class, it follows: 
\begin{eqnarray}
&&\left\| A_{\mathbf{q}}W_{\mathbf{q}}A_{\mathbf{q}}\right\| _{1} \\
&\leqslant &\frac{1}{2}\left\| \cosh \left[ \tfrac{\beta }{2}\left(
H_{n}^{\left( K,\mathbf{q}\right) }-\mu _{n}\right) \right] ^{-1}\left( -%
\tfrac{1}{2}\Delta _{\vec{\theta}_{\mathbf{q}}}+a\right) ^{4}\right\|  \notag
\\
&&\times \left\| e^{-\frac{\beta }{2}H_{n}^{\left( K,\mathbf{q}\right)
}}-e^{-\frac{\beta }{2}(-\frac{1}{2}\Delta _{\vec{\theta}_{\mathbf{q}%
}})}\right\| _{1}  \notag \\
&&+\frac{1}{2}\left\| \left( -\tfrac{1}{2}\Delta _{\vec{\theta}_{\mathbf{q}%
}}+a\right) ^{4}e^{\frac{\beta }{4}\Delta _{\vec{\theta}_{\mathbf{q}%
}}}\right\|  \notag \\
&&\times \left\| \cosh \left[ \tfrac{\beta }{2}\left( H_{n}^{\left( K,%
\mathbf{q}\right) }-\mu _{n}\right) \right] ^{-1}-\cosh \left[ \tfrac{\beta 
}{2}\left( -\tfrac{1}{2}\Delta _{\vec{\theta}_{\mathbf{q}}}-\mu _{n}\right) %
\right] ^{-1}\right\| _{1}\text{.}  \notag
\end{eqnarray}
Using Proposition 5, we can continue: 
\begin{eqnarray}
&&\left\| A_{\mathbf{q}}W_{\mathbf{q}}A_{\mathbf{q}}\right\| _{1} \\
&\leqslant &\dfrac{\lambda }{2}\left\{ \frac{\gamma _{F_{1}}}{\left(
1-\lambda \gamma _{a}\right) ^{4}}\sup_{x\in \lbrack \epsilon _{0},\infty
)}\left( x+a\right) ^{4}\cosh \left[ \tfrac{\beta }{2}\left( x-\mu
_{n}\right) \right] ^{-1}\right.  \notag \\
&&\left. +\gamma _{F_{2}}\sup_{x\in \lbrack 0,\infty )}\left( x+a\right)
^{4}e^{-\frac{\beta }{2}\left( x-\mu _{n}\right) }\right\} \left\| V^{\left(
K\right) }\left[ n\right] \right\| _{L^{1}\left( cell\right) }\text{,} 
\notag
\end{eqnarray}
where $F_{1}\left( z\right) =e^{-\frac{\beta }{2}z}$ and $F_{2}\left(
z\right) =\cosh \left[ \tfrac{\beta }{2}\left( z-\mu _{n}\right) \right]
^{-1}$. Moreover, 
\begin{eqnarray*}
&&\left\| V^{\left( K\right) }\left[ n\right] \right\| _{L^{1}\left(
cell\right) } \\
&\leqslant &2N_{0}L^{H}+v_{cell}^{1-1/p}\left\| v_{xc}\left( n\right)
\right\| _{L^{p}\left( cell\right) } \\
&\leqslant &2N_{0}L^{H}+N_{0}^{1/p}v_{cell}^{1-1/p}\sup_{t\in
R_{+}}t^{-1/p}v_{xc}\left( t\right) \text{.}
\end{eqnarray*}
Because $\mu _{n}\in \left[ \mu _{m},\mu _{M}\right] $, we can conclude that 
\begin{equation}
\left\| A_{\mathbf{q}}W_{\mathbf{q}}A_{\mathbf{q}}\right\| \leqslant \left\|
A_{\mathbf{q}}W_{\mathbf{q}}A_{\mathbf{q}}\right\| _{1}\leqslant ct.\lambda 
\text{,}
\end{equation}
where $ct.$ is independent of $\mathbf{q}$ or $K.$ Consequently, $\left\| W_{%
\mathbf{q}}\right\| _{L^{\infty }\left( cell\right) }\leqslant ct.\lambda $,
where $ct.$ is again independent of $\mathbf{q}$ or $K$. Then the
affirmation follows from 
\begin{eqnarray}
&&\left\| T_{K}\left[ n\right] -\left( 1+e^{\beta (-\frac{1}{2}\Delta -\mu
_{n})}\right) ^{-1}\left( \vec{x},\vec{x}\right) \right\| _{L^{\infty
}\left( \mathcal{T}\right) } \\
&=&\left\| K^{-3}\sum\nolimits_{\mathbf{q}\in \Lambda _{K}}W_{\mathbf{q}%
}\left( \vec{x},\vec{x}\right) \right\| _{L^{\infty }\left( cell\right)
}\leqslant ct.\lambda  \notag
\end{eqnarray}
and 
\begin{eqnarray}
&&\left\| \Phi _{FD}\left( -\tfrac{1}{2}\Delta -\mu _{n}\right) \left( \vec{x%
},\vec{x}\right) \right\| _{L^{\infty }\left( \mathcal{T}\right) } \\
&&-\left\| T_{K}\left[ n\right] -\Phi _{FD}\left( -\tfrac{1}{2}\Delta -\mu
_{n}\right) \left( \vec{x},\vec{x}\right) \right\| _{L^{\infty }\left( 
\mathcal{T}\right) }  \notag \\
&\leqslant &\left\| T_{K}\left[ n\right] \right\| _{L^{\infty }}\leqslant
\left\| \Phi _{FD}\left( -\tfrac{1}{2}\Delta -\mu _{n}\right) \left( \vec{x},%
\vec{x}\right) \right\| _{L^{\infty }\left( \mathcal{T}\right) }  \notag \\
&&+\left\| T_{K}\left[ n\right] -\Phi _{FD}\left( -\tfrac{1}{2}\Delta -\mu
_{n}\right) \left( \vec{x},\vec{x}\right) \right\| _{L^{\infty }\left( 
\mathcal{T}\right) }\text{,}  \notag
\end{eqnarray}
by observing that 
\begin{equation}
n_{\min }\leqslant \left\| \Phi _{FD}\left( -\tfrac{1}{2}\Delta -\mu
_{n}\right) \left( \vec{x},\vec{x}\right) \right\| _{L^{\infty }\left( 
\mathcal{T}\right) }\leqslant n_{\max }\text{.}\blacksquare
\end{equation}

Let us now return to the last condition (\textit{C}2) which remains to
verified. We define the set $B$ as a the strip in $L^{\infty }\left(
cell\right) $: 
\begin{equation}
B=\left\{ n\in L^{\infty }\left( cell\right) \text{, }n_{\min }-\varepsilon
\leqslant n\leqslant n_{\max }+\varepsilon \text{, }a.e.\right\} \text{,}
\end{equation}
where $\varepsilon $ is a positive constant such that $n_{\min }>\varepsilon 
$. Observing that $v_{xc}^{\prime }\left( t\right) $ is bounded over $I=%
\left[ n_{\min }-\varepsilon ,n_{\max }+\varepsilon \right] $, it follows: 
\begin{eqnarray}
&&\left\| v_{xc}\left( n_{1}\right) -v_{xc}\left( n_{2}\right) \right\|
_{L^{1}\left( cell\right) }  \notag \\
&=&\int_{cell}\left| \int_{n_{1}\left( \vec{x}\right) }^{n_{2}\left( \vec{x}%
\right) }\frac{dv_{xc}\left( t\right) }{dt}\right| d\vec{x} \\
&\leqslant &\int_{cell}\sup_{t\in I}\left| v_{xc}^{\prime }\left( t\right)
\right| \left| n_{1}\left( \vec{x}\right) -n_{2}\left( \vec{x}\right)
\right| d\vec{x}  \notag \\
&\leqslant &\sup_{t\in I}\left| v_{xc}^{\prime }\left( t\right) \right|
\left\| n_{1}-n_{2}\right\| _{L^{1}\left( cell\right) }\text{.}  \notag
\end{eqnarray}
where we omitted a set of zero measure where $n\left( \vec{x}\right) $ can
have values which are not in the interval $I$. For $\lambda $ smaller than a
certain value, independent of $K$, it follows from Theorem 14 that $T_{K}%
\left[ S_{per}^{K}\right] \subset B_{K}$ which completes our analysis of the
local density approximation.

\textbf{Acknowledgement.} This work was supported by the Robert A. Welch
foundation under grant C-1222.


\begin{thebibliography}{99}
\bibitem{Eschrig}  H. Eschrig, \emph{The Fundamentals of Density Functional
Theory} (Teubner  Verlagsgesellschaft, Stuttgart, 1996).

\bibitem{Hohenberg64}  D. Hohenberg and W. Kohn, Phys. Rev. \textbf{136},
B864 (1964).

\bibitem{Mermin65}  N.~D. Mermin, Phys. Rev. \textbf{A137}, 1441 (1965).

\bibitem{KohnSham65}  W. Kohn and L.~J. Sham, Phys. Rev. \textbf{140}, 1133
(1965).

\bibitem{Stoitsov88}  M.~V. Stoitsov and I.~Z. Petkov, Ann. Physics \textbf{%
184}, 121 (1988).

\bibitem{Bokanowski00}  I.~S. O.~Bokanowski and H. Zidani, Nonlinear Anal. 
\textbf{41}, 33 (2000).

\bibitem{Kaiser99}  H.~C. Kaiser and J. Rehberg, Z. Angew. Math. Phys. 
\textbf{50}, 423 (1999).

\bibitem{Lions98}  I. Catto, C.~L. Bris, and P. Lions, \emph{The
Mathematical Theory of  Thermodynamics Limits: Thomas-Fermi Type Models}
(Oxford University Press,  Oxford, 1998).

\bibitem{Catto01}  I. Catto, C.~L. Bris, and P.~L. Lions, Ann. Inst. H.
Poincare \textbf{18}, 687  (2001).

\bibitem{ProdanHI}  E. Prodan and P. Nordlander, J. Math. Phys. \textbf{42},
3390 (2001).

\bibitem{ProdanHII}  E. Prodan and P. Nordlander, J. Math. Phys. \textbf{42}%
, 3407 (2001).

\bibitem{ProdanHIII}  E. Prodan and P. Nordlander, J. Math. Phys. \textbf{42}%
, 3424 (2001).

\bibitem{Lieb94}  V. Bach, E.~H. Lieb, M. Loss, and J.~P. Solovej, Phys.
Rev. Lett. \textbf{72},  2981 (1994).

\bibitem{Lieb85}  E.~H. Lieb, in \emph{Density Functional Methods in Physics}%
, edited by R.  Dreizler and J.~D. Providencia (Plenum Press, New York,
1985), pp.\ 31--80.

\bibitem{Mahan90book}  G.~D. Mahan, \emph{Many-Particle Physics} (Plenum,
New York, 1990).

\bibitem{Lieb77}  E. Lieb and B. Simon, Commun. Math. Phys. \textbf{53}, 185
(1977).

\bibitem{Lions87}  P. Lions, Commun. Math. Phys. \textbf{109}, 33 (1987).

\bibitem{SimonTr}  B. Simon, \emph{Trace Ideals and Their Applications}
(Cambridge University  Press, New York, 1979).

\bibitem{SimonIV}  M. Reed and B. Simon, \emph{Methods of Modern
Mathematical Physics} (Academic  Press, New York, 1978), Vol.~IV.

\bibitem{SimonI}  M. Reed and B. Simon, \emph{Methods of Modern Mathematical
Physics} (Academic  Press, New York, 1972), Vol.~I.

\bibitem{Perdew92}  J. Perdew and Y. Wang, Phys. Rev. \textbf{B45}, 13244
(1992).
\end{thebibliography}
\end{document}